\documentclass[twocolumn,preprintnumbers,amsmath,nofootinbib,amssymb]{revtex4}
\usepackage{amssymb}
\usepackage{latexsym}
\usepackage{color}
\usepackage{enumerate}
\usepackage{graphicx}
\usepackage{epsfig}
\usepackage{hyperref}
\usepackage{color}

\def\be{\begin{equation}}
\def\ee{\end{equation}}
\def\bea{\begin{eqnarray}}
\def\eea{\end{eqnarray}}
\begin{document}
\title{The third law of thermodynamics and black holes}
\author{H. Moradpour$^1$\footnote{hn.moradpour@maragheh.ac.ir}, A. H. Ziaie$^1$\footnote{ah.ziaie@maragheh.ac.ir}, Iarley P. Lobo$^2$\footnote{iarley\_lobo@fisica.ufpb.br}, J. P. Morais Gra\c{c}a$^3$\footnote{jpmorais@gmail.com}, U. K. Sharma$^4$\footnote{sharma.umesh@gla.ac.in}, A. Sayahian Jahromi$^5$}
\address{$^1$ Research Institute for Astronomy and Astrophysics of Maragha
(RIAAM), University of Maragheh, P.O. Box 55136-553, Maragheh,
Iran\\
$^2$ Department of Chemistry and Physics, Federal University of Para\'iba, Rodovia BR 079 - Km 12, 58397-000 Areia-PB,  Brazil\\
$^3$ Instituto de F\'{i}sica, Universidade Federal do Rio de Janeiro, 21.941-972 - Rio de Janeiro-RJ-Brazil\\
$^4$ Department of Mathematics, Institute of Applied Sciences and
Humanities, GLA University,
    Mathura-281406, Uttar Pradesh, India\\
$^5$ Zarghan Branch, Islamic Azad University, Zarghan, Iran}
\date{\today}
\begin{abstract}
Working in the framework of generalized statistics, the problem of
availability of the third law of thermodynamics in the black hole
physics is studied by focusing on Schwarzschild black hole which
easily and clearly exposes the violation of this law in the common
approach based on Bekenstein entropy. Additionally, it is
addressed that some inconsistencies between the predictions of
quantum field theory and those of thermodynamics on the black hole
temperature may be reconciled by using the thermodynamics laws in
order to broaden energy definition. It claims that thermodynamics
should be employed as a powerful tool in looking for more
comprehensive energy definitions in high energy physics, still
mysterious.
\end{abstract}
 \pacs{....}

\maketitle

\section{Introduction}

The third law of thermodynamics states that the entropy of a
system should approach a constant value ($C$) at absolute zero
temperature, or equally $S(T\rightarrow0)\rightarrow C$.
Bekenstein entropy ($S_B$) is proportional to the horizon area
$(A=4\pi r_h^2)$, and correspondingly, for a Schwarzschild black
hole of mass $M(\equiv E)$ and Hawking temperature $T_H$ for which
$r_h=2M$ and $T=\frac{\partial M}{\partial S}=\frac{1}{8\pi
M}=T_H$, we have

\begin{eqnarray}\label{1}
S_B=\frac{A}{4}=4\pi M^2=\frac{1}{16\pi T_H^2},
\end{eqnarray}

\noindent in the units of $c=\hbar=k_B=G=1$, where $k_B$ denotes
the Boltzmann constant. It clearly indicates that the third law of
thermodynamics is not satisfied or briefly,
$T_H\rightarrow0(\parallel M\rightarrow\infty)\Rightarrow
S_B\rightarrow\infty$. Although we considered Schwarzschild black
hole, the behavior of
$S_B\big(T(M\rightarrow\infty)\rightarrow0\big)\rightarrow\infty$
is common in other black holes such as Kerr-Newman and
Reissner-Nordstr\"{o}m metrics \cite{1,01,2,3,4}.

$S_B$ is non-extensive, a trait which reminds the generalized
statistics such as Tsallis statistics \cite{revT,masi} including
entropies which are not extensive. Indeed, as gravitational
systems include long-range interaction (gravity), it is proposed
that the Boltzmann entropy (leading to Eq.~(\ref{1})) should be
replaced with generalized entropies which leads to substantial
consequences in both gravitational and cosmological setups (see
for example
Refs.~\cite{tsallis,refgerg,gerg,non13,nonK,EPJC,KHDE,sadeghi,mesri2}
their references and citations). It also seems that there is a
connection between deviation from the Boltzmann statistics and the
quantum features of gravity \cite{epl,homa,mesri,barrow,mesri2},
and consequently, the generalized and loop quantum gravity
entropies can be classified as the subclasses of a general entropy
\cite{gen}.

Our first aim is to study the possibility of satisfying the third
law by employing some new entropies proposed that provide
respectable solutions in cosmological and gravitational setups.
Hereof and in subsequent sections, we respectively study the
problem by considering Tsallis and Cirto entropy, Tsallis entropy
and Kaniadakis entropy. Throughout the survey, we also address a
thermodynamic energy definition for a black hole of mass $M$,
corresponding to each entropy, which admits Hawking temperature.
The black hole remnant and its decay time in mentioned entropy
formalisms have been studied in fifth section. The last section is
also devoted to summary.

\section{Tsallis and Cirto entropy and the third law of
thermodynamics}\label{TC00}

Motivated by the non-extensivity of $S_B$, and also the long-range
nature of gravity, Tsallis and Cirto \cite{tsallis} introduce a
new entropy for black holes as

\begin{eqnarray}\label{2}
S_T=\gamma A^\delta,
\end{eqnarray}

\noindent where $\gamma$ and $\delta$ are two unknown free
constants evaluated by other parts of physics or observations. It
is also useful to note here that this form of entropy is also
supported in the framework of quantum gravity \cite{barrow}. It
means that two different approaches and motivations lead to the
same result which increases the credit of this new proposal for
the black hole entropy. Moreover, the equality of results can be
considered as the sign of a deep connection between
non-extensivity and quantum gravity helping us build a relation
between their parameters, a result noted in Refs.~\cite{epl,homa}.
Considering $r_h=2M$ and $T=\frac{\partial M}{\partial
S_T}=\frac{1}{2\delta\gamma(16\pi)^\delta M^{2\delta-1}}$, one
easily finds that the third law of thermodynamics is met whenever
$0<\delta<\frac{1}{2}$, and in summary, $S_T\rightarrow0\parallel
M\rightarrow0\parallel T\rightarrow0$.

Now, let us employ Hawking temperature ($T_H=\frac{1}{8\pi M}$)
instead of $T=\frac{\partial M}{\partial S_T}$ which leads to
$S_T\propto T^{-2\delta}$ meaning that the third law is fulfilled
only if $\delta<0$ and for this case, we briefly have
$M\rightarrow\infty\parallel S_T,T\rightarrow0$. In Tsallis
statistics, an intrinsic temperature discrepancy between real
temperature and the temperature obtained by thermodynamic relation
may emerge depending on the expectation values definition
(averaging methods) used in obtaining quantities such as energy
\cite{kol}, and only, having in hand the system temperature, one
can decide on true temperature \cite{kol}. Therefore, the above
obtained temperature discrepancy may be more understandable by
bearing in mind the intrinsic temperature discrepancy of Tsallis
statistics. Consequently, since $\delta$ is a free parameter
estimated from observations \cite{revT,masi} or probably other
parts of physics \cite{epl,homa}, we cannot go further in choosing
one of these temperatures, and thus the corresponding
thermodynamics, unless we enclasp detailed observations, data and
info on black holes.

Of course, a way to reconcile the above inconsistency between
temperatures is to redefine energy. In both cases above, we
assumed $E=M$, while if we assume $T=T_H=\frac{\partial
E}{\partial S}$, and use Eq.~(\ref{2}), then we reach

\begin{eqnarray}
E_T=\int_0^M\frac{1}{8\pi m}\frac{\partial S_T}{\partial m}dm,
\end{eqnarray}

\noindent finally leading to

\begin{eqnarray}\label{3}
E_T=4\gamma\delta\frac{(4\pi)^{\delta-1}}{2\delta-1}M^{2\delta-1},
\end{eqnarray}

\noindent as the energy of a Schwarzschild black hole of mass $M$
in Tsallis formalism which recovers $E=M$ by inserting $\delta=1$
and $\gamma=\frac{1}{4}$ (the Bekenstein limit). It is also
obvious that $E_T$ is positive if $\delta>\frac{1}{2}$ or
$\delta<0$. For the $\delta>\frac{1}{2}$ case, the third law is
not satisfied and we face a situation like to what we obtained in
the case of Bekenstein entropy (i. e, $T_H\rightarrow0\parallel
S_B,M,E_T\rightarrow\infty$). The third law is met for $\delta<0$
and in parallel $E_T\rightarrow0$ or briefly,
$E_T,S_T,T\rightarrow0\parallel M\rightarrow\infty$.

Finally, we think that since the Hawking temperature is also
supported by other parts of physics such as quantum field theory
in curved spacetime and so on \cite{haw,bek,thes}, and as there is
not any common agreement on the energy definition in high energy
physics \cite{energy,energy1}, it is probably reasonable to rely
on this approach compared to the two previously mentioned cases.
\textit{The latter means that thermodynamics may be used to find a
more proper energy definition in high energy physics.}

In summary, we obtained $3$ cases recaped as

\begin{eqnarray}\label{8}
&&\!\!\!\!i)\ E=M,\ T=\frac{\partial E}{\partial S_T}=\frac{1}{2\delta\gamma(16\pi)^\delta M^{2\delta-1}}\neq T_H,\nonumber\\
&&\!\!\!\!ii)\ E=M,\ T=T_H=\frac{1}{8\pi M}\neq\frac{\partial E}{\partial S_T},\\\nonumber
&&\!\!\!\!\!\!\!\!iii)\ E_T=4\gamma\delta\frac{(4\pi)^{\delta-1}}{2\delta-1}M^{2\delta-1},\ T=T_H=\frac{1}{8\pi M}=\frac{\partial E_T}{\partial S_T},
\end{eqnarray}

\noindent where the third law is satisfied for the first case when
$0<\delta<\frac{1}{2}$, and for the remaining cases when
$\delta<0$.

\section{Tsallis entropy and the third law}

Recently, focusing on the relation between Tsallis and Boltzmann
statistics, a new entropy has been derived for black holes as
\cite{KHDE}

\begin{eqnarray}\label{T1}
&&S_q=\frac{1}{1-q}[\exp\big((1-q)S_{B}\big)-1]\nonumber\\
&&=\frac{2\exp(\frac{(1-q) S_{B}}{2})}{(1-q)}\sinh\left(\frac{(1-q)
S_{B}}{2}\right),
\end{eqnarray}

\noindent where $q$ is a free unknown parameter and this result is
also confirmed by calculating the Tsallis entropy content of black
holes in the framework of quantum gravity \cite{mesri,KHDE}.
Following the recipe of previous section, if we assume $E=M$ then
we have

\begin{eqnarray}\label{T2}
T=\frac{\partial E}{\partial S_q}=T_H\exp\big((q-1)4\pi M^2\big),
\end{eqnarray}

\noindent which recovers $T_H$ whenever $q=1$ (the Bekenstein
limit of~(\ref{T1}) \cite{KHDE}). Briefly, $T\rightarrow0$ only if
$q<1$, and in this manner $M, S_q\rightarrow\infty$ meaning that
the third law is not satisfied. On the other hand, if we assume
$T=T_H$ (case $ii$), then we see that the third law is satisfied
only if $q>1$, or briefly $M\rightarrow\infty\Rightarrow
S_q(T\rightarrow0)\rightarrow0$. For the case $iii$, where
$T=T_H=\frac{1}{8\pi M}=\frac{\partial E_q}{\partial S_q}$, we
reach

\begin{eqnarray}\label{T3}
E_q=\int_0^M\exp\big((1-q)4\pi m^2\big)dm,
\end{eqnarray}

\noindent as the energy content of a black hole of mass $M$.
Clearly, the third law is again satisfied only if $q>1$ and
moreover, $E=\int_0^Mdm=M$ is recovered at the limit of
$q\rightarrow1$. The above integration can be performed with the
solution given as

\begin{eqnarray}\label{T4}
E_q=\frac{{\rm
erf}\left(2\sqrt{\pi}M\sqrt{q-1}\right)}{4\sqrt{q-1}},
\end{eqnarray}

\noindent in which $\rm erf(x)$ denotes the error function
\cite{er}.

\section{Kaniadakis entropy and the third law}

Kaniadakis entropy of a black hole is also reported as \cite{KHDE}

\begin{eqnarray}\label{K1}
S_\kappa=\frac{1}{\kappa}\sinh\big(\kappa S_{B}\big),
\end{eqnarray}

\noindent where $\kappa$ is an unknown parameter evaluated by
observations and probably, the other parts of physics \cite{KHDE}.
Here, simple calculations lead to

\begin{eqnarray}\label{K2}
T=\frac{\partial E}{\partial S_\kappa}=\frac{T_H}{\cosh(\kappa S_B)},
\end{eqnarray}

\noindent for the $i$-th case. The result indicates that,
independent of the value of $\kappa$, the third law is not
satisfied ($S_\kappa\rightarrow\infty\parallel
M\rightarrow\infty\parallel T\rightarrow0$). For the second case
($T=T_H, E=M$), we can write

\begin{eqnarray}\label{K3}
S_\kappa=\frac{1}{\kappa}\sinh\left(\frac{\kappa}{16\pi T^2}\right),
\end{eqnarray}

\noindent which shows the third law is satisfied only if
$\kappa<0$. If $\kappa<0$, then the third law is also met for case
$iii$, where $T=T_H$ and energy content of black hole is
obtainable by using

\begin{eqnarray}\label{K4}
E_\kappa=\int_0^M\cosh(\kappa4\pi m^2)dm,
\end{eqnarray}

\noindent which recovers the $E=\int_0^Mdm=M$ results at the limit
of $\kappa=0$. The solution to the above integral is also given as

\begin{eqnarray}\label{K5}
E_\kappa=\frac{1}{8\sqrt{\kappa}}\left[{\rm erf}\left(2\sqrt{\kappa\pi}M\right)+{\rm erfi}\left(2\sqrt{\kappa\pi}M\right)\right],
\end{eqnarray}
where
\begin{equation}
{\rm erfi}(x)=-i{\rm erf}(ix).
\end{equation}

\begin{figure}
    \begin{center}
        \includegraphics[scale=0.45]{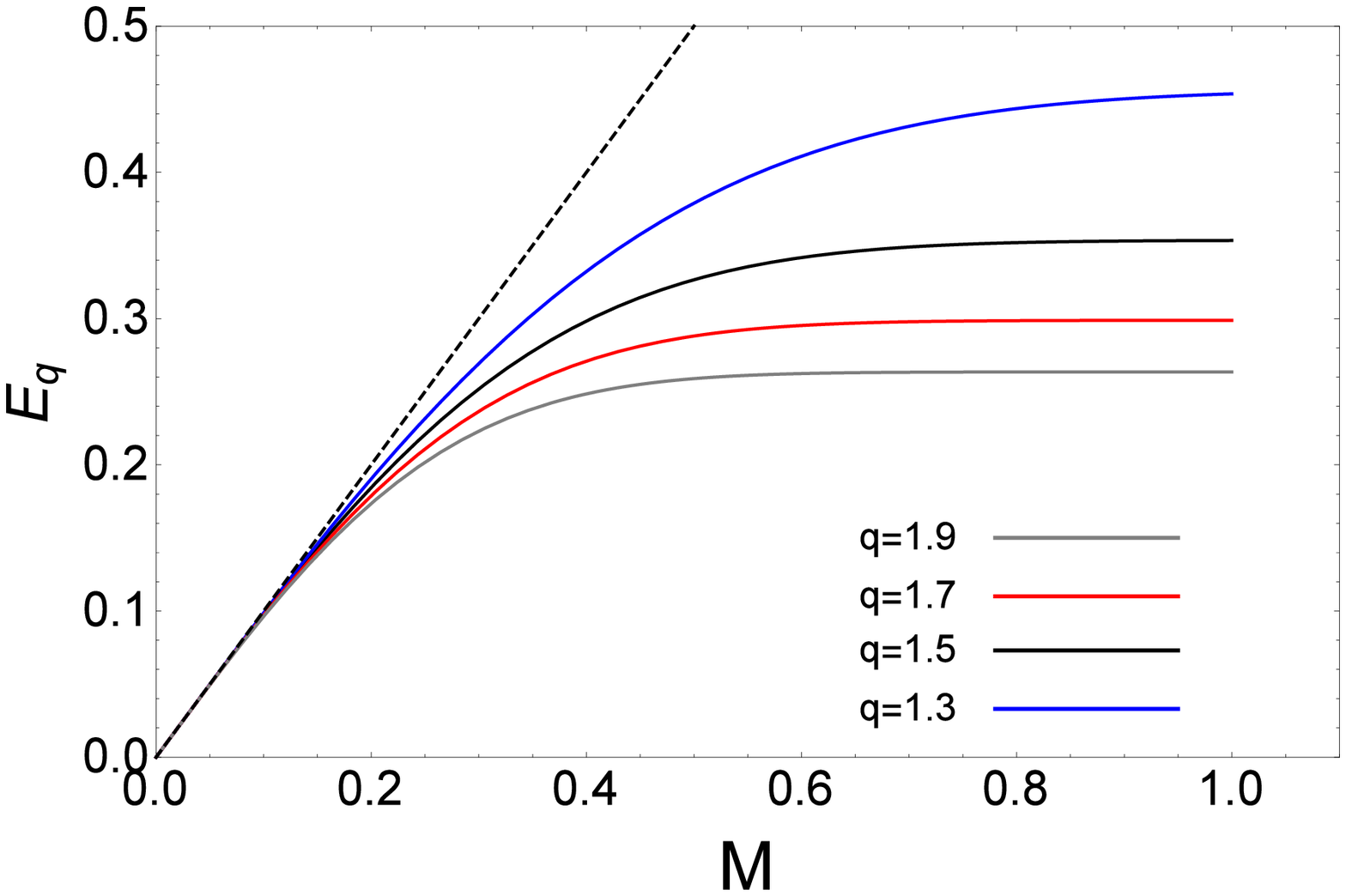}
        \includegraphics[scale=0.45]{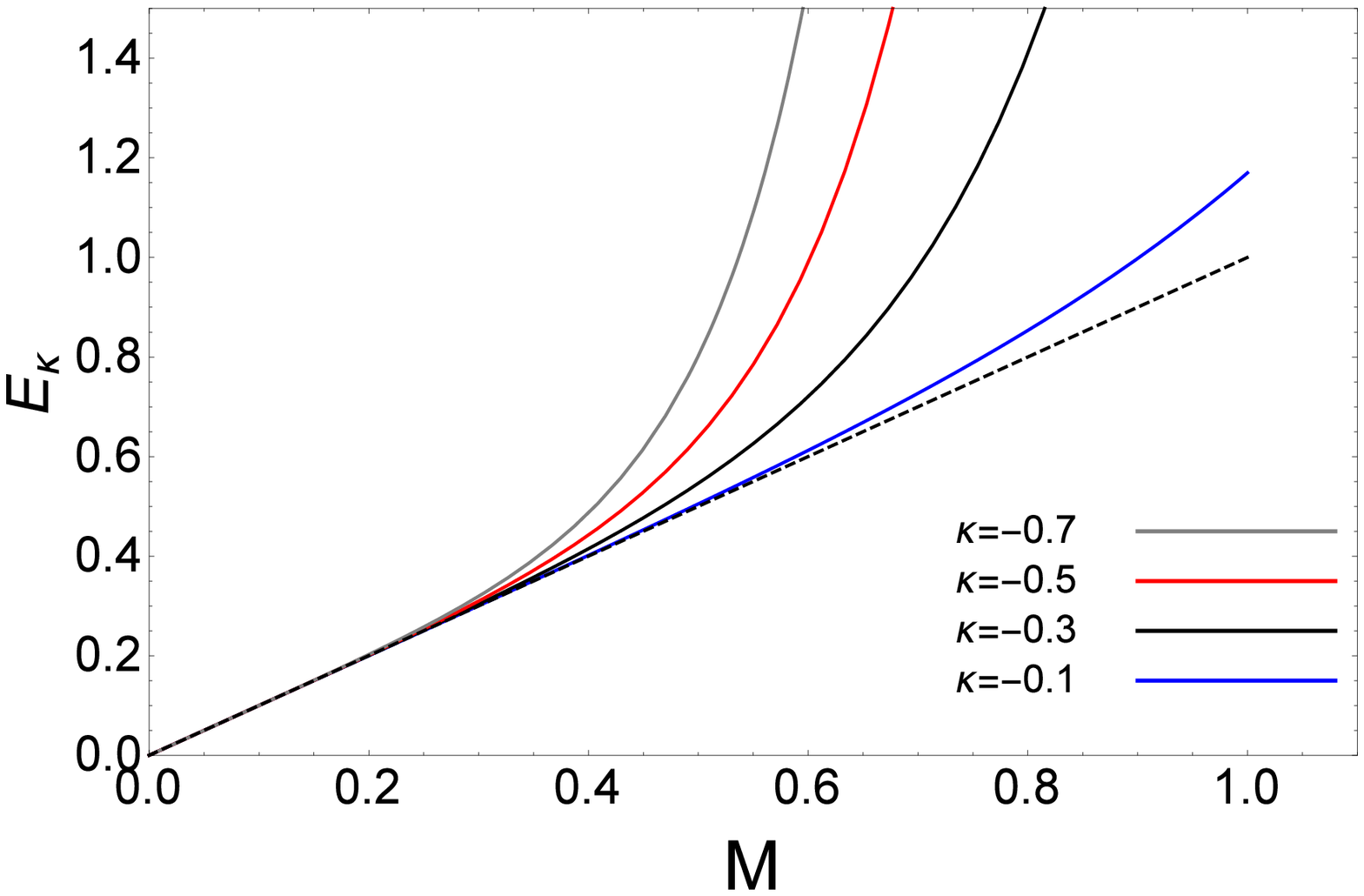}
        \caption{Upper panel: The behavior of energy content of the Schwarzschild black hole as a function of its mass for different values of $q$ parameter. Lower panel: The behavior of energy content of the Schwarzschild black hole as a function of its mass for different values of $\kappa$ parameter. The dashed curve also shows the Schwarzschild black hole of $E=M$.}\label{fig1}
    \end{center}
\end{figure}

In Fig.~(\ref{fig1}), $E_\kappa$ and $E_q$ are plotted for
different values of $q$ and $\kappa$, the $E=M$ case has also been
depicted to have a comparison. It is worthwhile to mention that,
as it is obvious from this figure, there is an asymptote for $E_q$
when $q>1$ as $E_q(M\gg1)\rightarrow\frac{1}{4\sqrt{q-1}}$.

\section{Black body radiation and black hole evaporation}

In the framework of common statistical mechanics based on Gibbs
entropy, the black hole evaporation is described by
Stefan-Boltzmann (SB) formula \cite{cav,ali}. As it is apparent
from Eq.~(\ref{1}), we have $S_B,M\rightarrow\infty$ when
$T\rightarrow0$ meaning that we face a catastrophe \cite{ali}.
This result can be summarized in the language of SB law as

\begin{eqnarray}\label{4}
\frac{dM}{dt}=-A\sigma T^4,
\end{eqnarray}

\noindent where the ordinary energy definition ($E=M$) is used and
$\sigma(=\frac{\pi^2}{60}$ in our units \cite{kanbb1,kelly})
denotes the SB constant \cite{ali}. Here, minus sign is also due
to the fact that Eq.~(\ref{4}) explains the amount of energy that
system loses it. It clearly shows that the decay rate
($\frac{dM}{dt}$) diverges as $M$ ($T$) approaches zero (infinity)
\cite{ali}. Another consequence of this law is our ability to find
the decay time $\tau$ as

\begin{eqnarray}\label{5}
\tau=-\frac{1}{\sigma}\int_{M}^0\frac{dm}{AT^4}=\frac{(8\pi)^3}{2\sigma}\frac{M^3}{3},
\end{eqnarray}

\noindent and therefore $\tau\sim M^3$ for a Schwarzschild black
hole of temperature $T_H$. Eq.~(\ref{5}) also indicates that
$\tau\rightarrow\infty$ when $M\rightarrow\infty$ while
$T,\frac{dM}{dt}\rightarrow0$ \cite{ali}.

\subsection{Tsallis and Cirto (TC) black hole}

Black body radiation has not been studied in the formalism of
Eq.~(\ref{2}), but as we mentioned, this entropy is also confirmed
by quantum gravity \cite{barrow}. The latter allows us to use the
black body radiation formula in the framework of quantum gravity
to study the black holes meeting Eq.~(\ref{2}). The modifications
of quantum gravity on the black body spectrum have recently been
studied by some authors in different bases
\cite{nozar,hus,cqg,lobo}. A common feature of quantum gravity
scenarios is generalized uncertainty principle (GUP) that
motivates us to focus on its modification on black body radiation.
Indeed, such a modification is also obtainable by using other
quantum gravity scenarios \cite{hus,cqg}.

In this regard, it is proposed that GUP may modify the black body spectrum as \cite{cqg}

\begin{eqnarray}\label{TC1}
\frac{dE}{dt}=-A\sigma\big[T^4 +\frac{15}{9}\alpha T^6\big],
\end{eqnarray}

\noindent where $\alpha$ is called the GUP parameter originated
from the quantum aspects of gravity, and there is an ongoing
debate on its value \cite{agha,cqg}. It is also obvious that
Eq.~(\ref{4}) is recovered whenever $\alpha\rightarrow0$ and
$E=M$. Calculations for three obtained cases lead to

\begin{eqnarray}\label{TC2}
\!\!\!\!\!\!\!\tau_i^{TC}&=&-\frac{(2\delta\gamma(16\pi)^\delta)^4}{16\pi\sigma}\int_{M}^0\frac{dm}{m^{6-8\delta}+\frac{15}{(6\delta\gamma(16\pi)^\delta)^2}\alpha m^{8-12\delta}},\nonumber\\
\!\!\!\!\!\!\!\tau_{ii}^{TC}&=&-\frac{1}{18\sigma}\Big[1536\pi^3 M^3-120\pi M\alpha\nonumber\\&+&5\sqrt{15}\alpha^{\frac{3}{2}}\tan^{-1}(\frac{8\sqrt{3}\pi M}{\sqrt{5\alpha}})\Big],\nonumber\\
\!\!\!\!\!\!\!\tau_{iii}^{TC}&=&-\frac{64\gamma\delta(4\pi)^{\delta+4}}{\sigma}\int_{M}^0\frac{m^{2+2\delta}dm}{(8\pi m)^2+\frac{15}{9}\alpha},
\end{eqnarray}

\noindent where hypergeometric functions are solutions to the
first and third cases. It is also useful to mention that, in all
cases, the integrands recover Eq.~(\ref{5}) at the corresponding
appropriate limit. Since the third law is satisfied for the first
case and other cases when $0<\delta<\frac{1}{2}$, and $\delta<0$,
respectively, we plot the obtained evaporation times for some
values of $\delta$ that fall into these intervals.

\begin{figure}
    \begin{center}
        \includegraphics[scale=0.44]{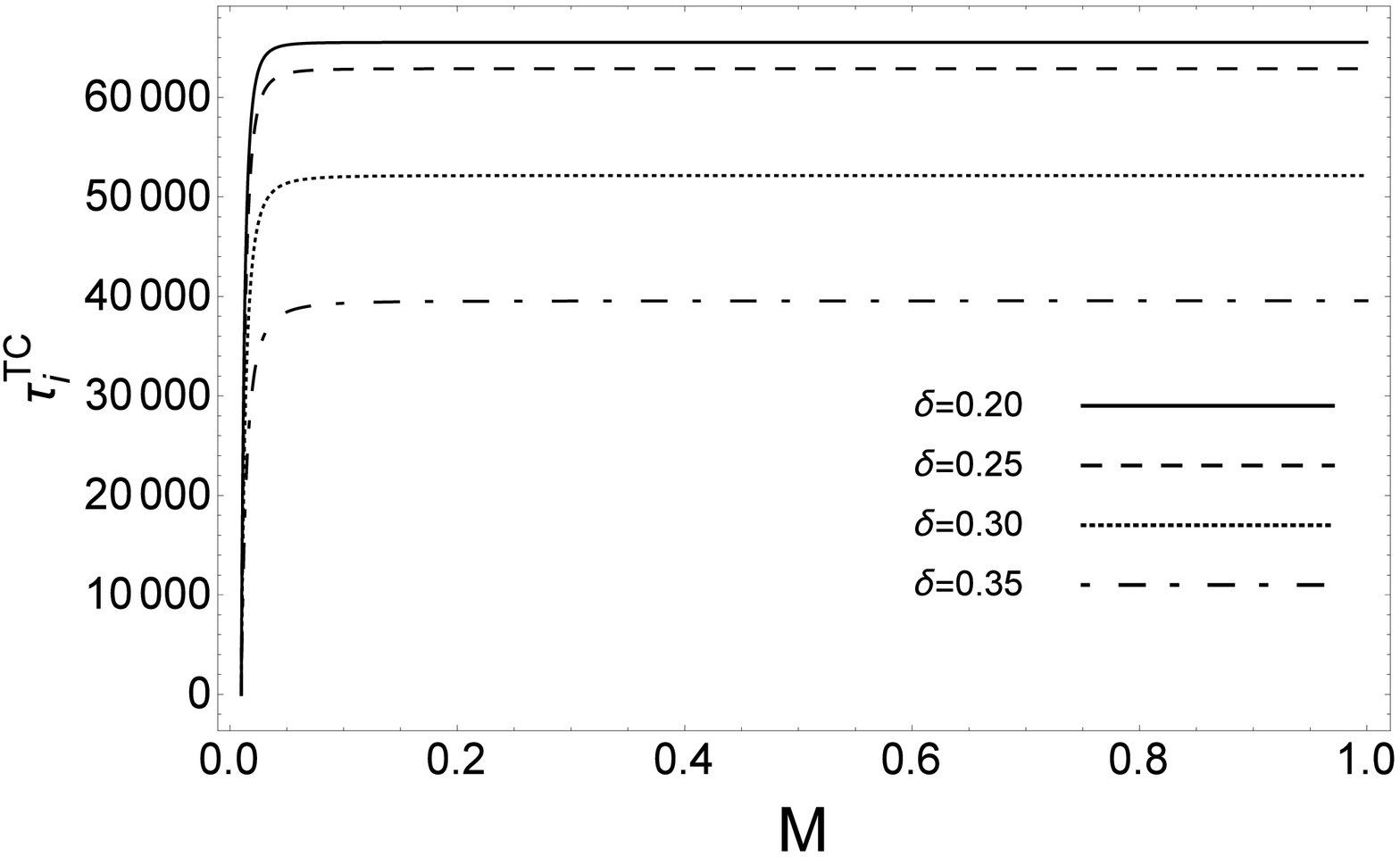}
        \includegraphics[scale=0.44]{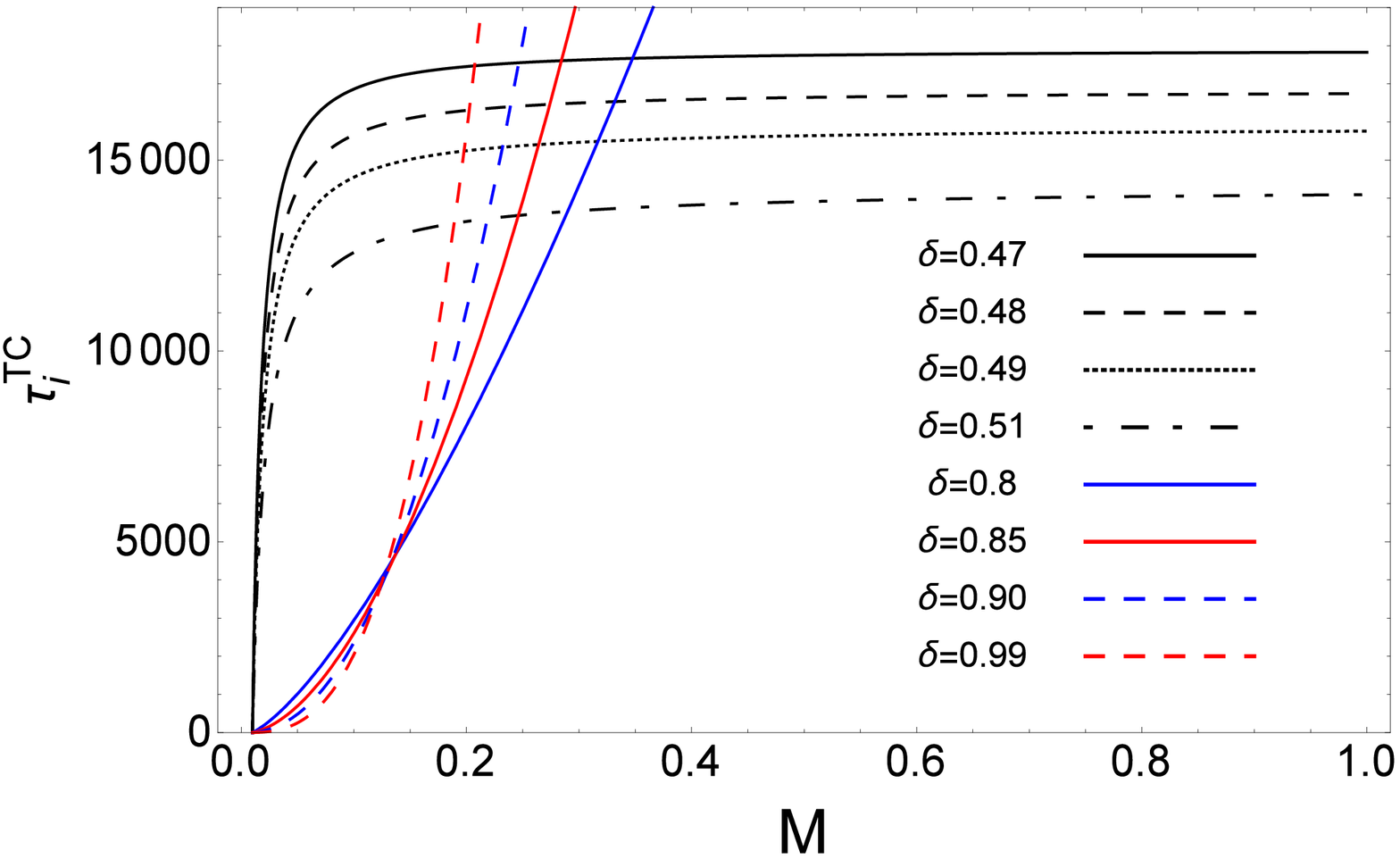}
        \includegraphics[scale=0.44]{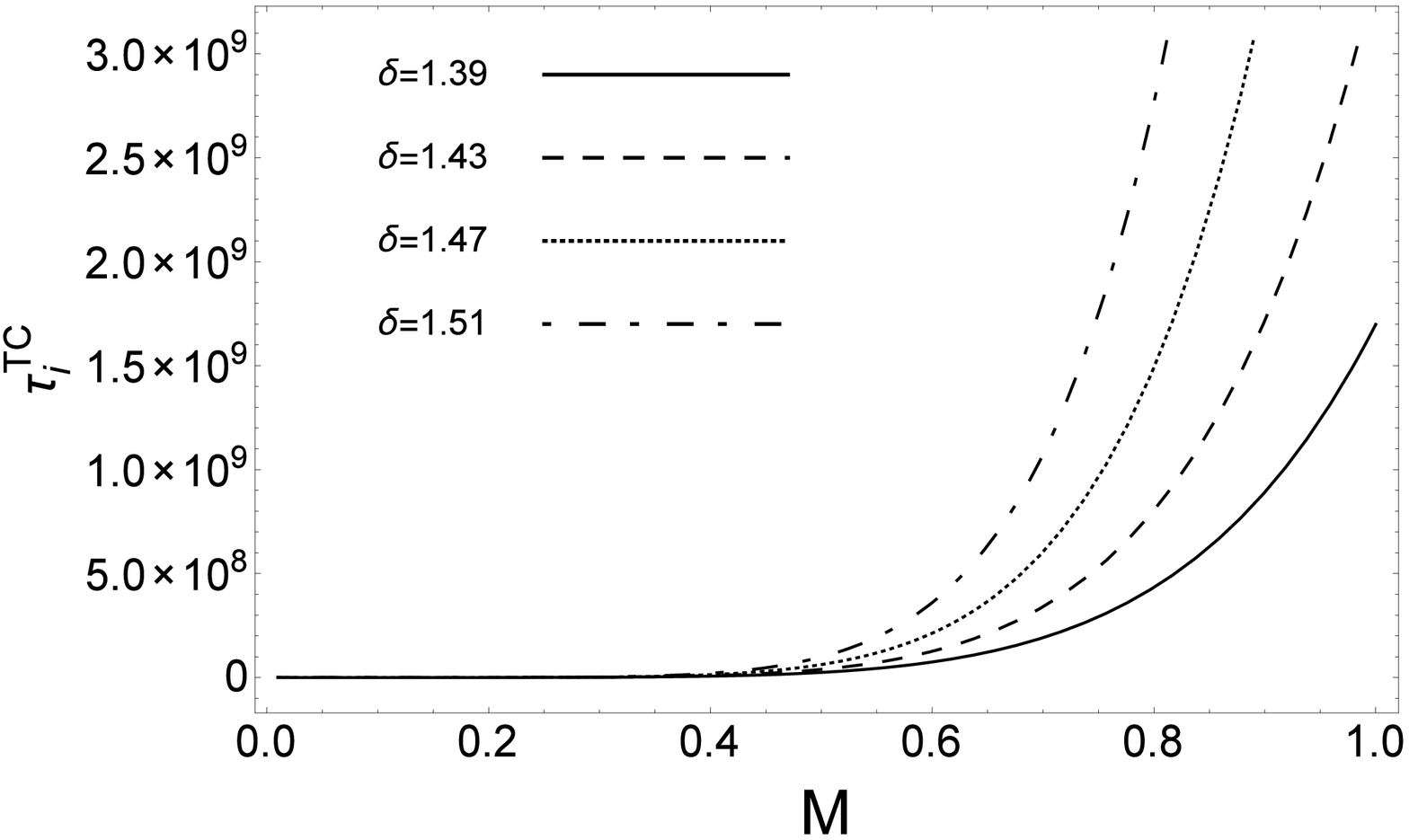}
        \caption{Plot of decay time for the first case, versus black hole mass for different values of parameter $\delta$. We have set $\sigma=\pi^2/60$, $\gamma=1$ and $\alpha=1$.}\label{fig2}
    \end{center}
\end{figure}

Figure (\ref{fig2}) shows the behavior of decay time against black
hole mass for different values of $\delta$ parameter. Each point
on the curves presents the time needed for a black hole of mass
$0<M\leq1$ to completely decay to zero mass. As we observe within
the upper panel, the decay time is finite and grows as the initial
black hole mass increases and asymptotically reaches a finite
value. Indeed, for $0<\delta<\frac{1}{2}$, it takes finite time
for a black hole to evaporate. The middle panel shows the behavior
of decay time for $\delta<1/2$ (family of black curves) and
$1/2<\delta<1$ (family of red and blue curves) where we observe
that there exists a critical value of this parameter
($\delta_{c}$) so that for $\delta<\delta_c$ the decay time is
finite and for $\delta>\delta_c$ the decay time diverges. In the
lower panel, we sketched the behavior of decay time for $\delta>1$
where it is seen that the decay time for a black hole of finite
mass grows unboundedly and diverges as the initial mass of black
hole increases. However, for these values of $\delta$ parameter
and those of middle panel, the third law of thermodynamics
($S(T\rightarrow0)\rightarrow 0$) is violated.

\begin{figure}
    \begin{center}
        \includegraphics[scale=0.44]{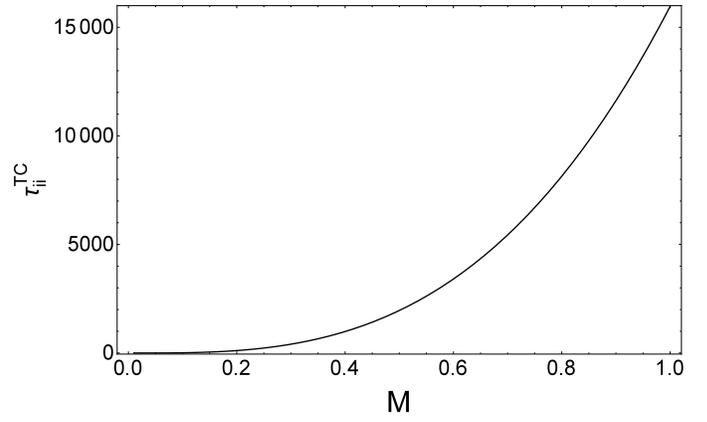}
                \caption{Plot of decay time for the second case, versus black hole mass. We have set $\sigma=\pi^2/60$ and $\alpha=1$. }\label{fig3}
    \end{center}
\end{figure}

\begin{figure}
    \begin{center}
        \includegraphics[scale=0.4]{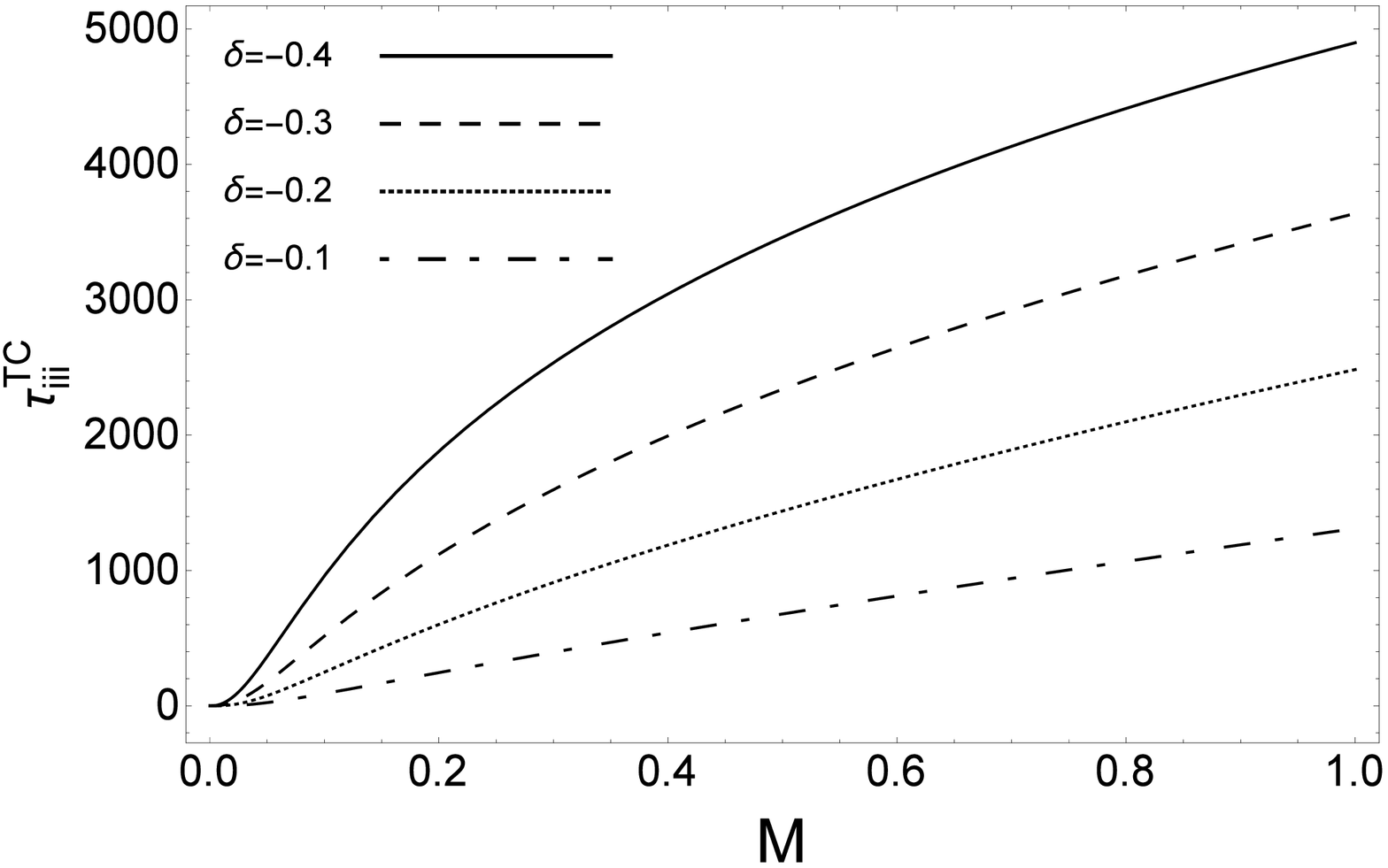}
        \includegraphics[scale=0.4]{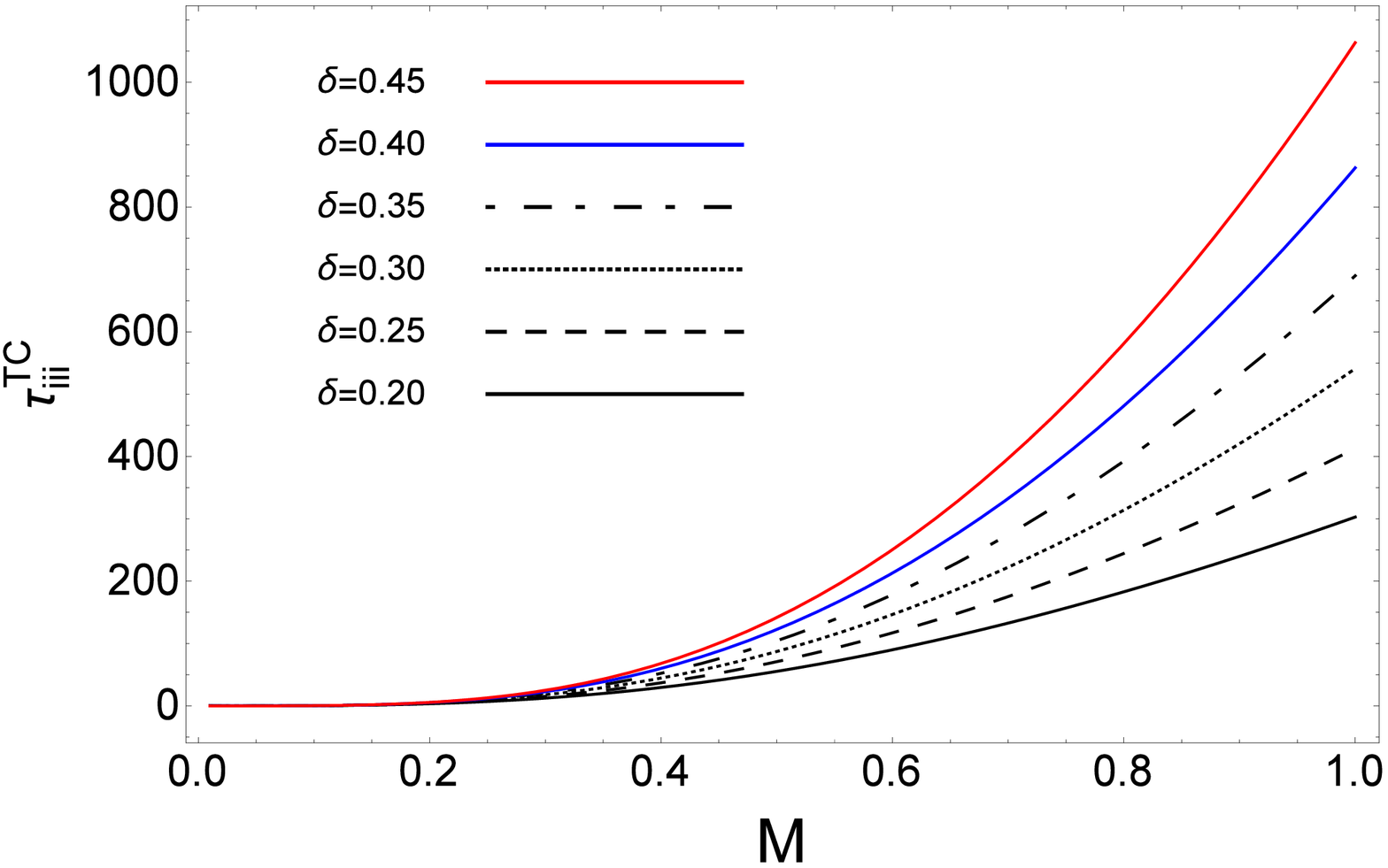}
        \caption{Plot of decay time for the third case, versus black hole mass. We have set $\sigma=\pi^2/60$, $\gamma=1$ and $\alpha=1$. }\label{fig4}
    \end{center}
\end{figure}

In Fig.(\ref{fig3}) we have plotted decay time of a black hole
against its mass for the second case, where we observe that the
larger the black hole mass the longer it takes for the black hole
to completely evaporate. The slope of $\tau^{TC}_{ii}$ curve
increases for larger values of black hole mass so that the decay
time grows unboundedly and diverges for massive and super massive
black holes. Finally, Fig. (\ref{fig4}) presents the decay time of
black hole for the third case. In the upper panel we observe that,
for $\delta<0$ for which the third law is respected, a black hole
with initial finite mass will completely evaporate at a finite
amount of time. The lower panel shows that for $\delta>0$ the
decay time is an increasing function of the black hole mass and
the heavier the initial black hole the longer it takes to
completely evaporate. However, from the viewpoint of the third
case, for this value of $\delta$ parameter the third law is not
respected. As yet there is no agreement on the numeric value of
$\alpha$ parameter we have considered the value of this parameter
to be unity~\cite{cqg,agha}. We further note that entropy is a
dimensionless quantity in each system of units where the Boltzmann
constant is unity.  Hence, from Eq. (\ref{2}) we can deduce that
$\gamma\propto C/\ell_{\rm Pl}^{2\delta}$ and as we work in the
system of units for which $\ell_{\rm Pl}=1$, $\gamma$ parameter is
a positive constant value which we have considered it to be unity.
\subsection{Tsallis black hole}

Although Eq.~(\ref{4}) is used in some previous works which
investigate the black hole thermodynamics in various non-extensive
statistics \cite{epl,refgerg,gerg,mesri2}, a comprehensive study
in the Tsallis framework should employs the Tsallis counterpart of
Eq.~(\ref{4}). The latter is a controversial issue
\cite{tbbr1,tbbr2,tbbr3,kol}, as different averaging methods are
useable and are employed in this statistics \cite{kol}. These
different methods have their own benefits and shortcomings, and
indeed, their correctness and accessibility situations are still
unsolved and need more attentions and observations \cite{kol}.
Here, motivated by the fact that

\begin{eqnarray}\label{6}
\frac{dE}{dt}=-A\sigma_q T^4
\end{eqnarray}

\noindent is obtained by different approaches \cite{tbbr1}, and is
also originated from the black body spectrum in Tsallis statistics
\cite{tbbs}, we focus on Eq.~(\ref{6}) as the alternative of
Eq.~(\ref{4}). Here, $\sigma_q$ is called the generalized SB
constant calculated as \cite{tbbr1}

\begin{eqnarray}\label{7}
&&\sigma_q=\frac{1}{\pi^2}\int_0^\infty\left[\frac{x^3}{\exp(x)-1}-\frac{1-q}{2}\frac{x^5\exp(x)}{(\exp(x)-1)^2}\right]dx\Rightarrow\nonumber\\
&&\sigma_q\simeq\sigma(1-6\cdot15(1-q))=\sigma(6\cdot15q-5\cdot15),
\end{eqnarray}

\noindent if the integration is numerically solved \cite{tbbr1}.
It is also obvious that $\sigma_q\rightarrow\sigma$ at the
appropriate limit of $q\rightarrow1$. For three obtained cases, we
have

\begin{eqnarray}\label{8}
&&\!\!\!\!\!\!\!\tau_i^{q}=-\frac{(8\pi)^3}{2\sigma_q}\int_M^0m^2\exp\big((1-q)4\pi m^2\big)dm,\nonumber\\
&&\!\!\!\!\!\!\!\tau_{ii}^{q}=\frac{(8\pi)^3}{2\sigma_q}\frac{M^3}{3},\\\nonumber
&&\!\!\!\!\!\!\!\tau_{iii}^{q}=\tau_i^{q}.
\end{eqnarray}

\begin{figure}
    \begin{center}
        \includegraphics[scale=0.4]{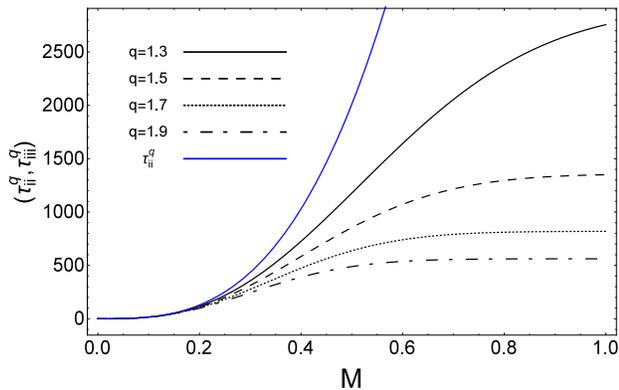}
        \caption{Plot of decay time for Tsallis black hole against its initial mass. We have set $\sigma=\pi^2/60$.}\label{fig5}
    \end{center}
\end{figure}

\noindent Since only the second and third cases satisfy the third
law only for $q>1$, in Fig. (\ref{fig5}), we plotted
$\tau_{ii}^{q}$ and $\tau_{iii}^{q}$ for different values of $q$
parameter (family of black curves). We therefore observe that the
decay time is finite and the lesser the value of $q$ parameter,
the longer it takes for a Tsallis black hole to completely
evaporate. The blue curve presents the behavior of $\tau_{ii}^{q}$
where we see that the decay time grows unboundedly as the initial
black hole mass increases.

%%%%%%%%%%%%%%%%%%%%%%%%%%%%%%%%%%%%%%

\begin{figure}
    \begin{center}
        \includegraphics[scale=0.4]{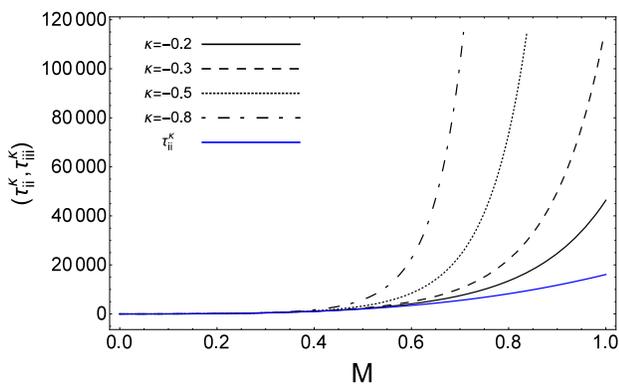}
        \caption{Plot of decay time for Kaniadakis black hole against its initial mass. We have set $\sigma=\pi^2/60$.}\label{fig6}
    \end{center}
\end{figure}

%%%%%%%%%%%%%%%%%%%%%%%%%%%%%%%%%%%%%%%%%%%%%%%%%%
\subsection{Kaniadakis black hole}

Black body spectrum in Kaniadakis statistics has recently been
studied \cite{kanbb1,kanbb2}, and it has been shown that
\cite{kanbb1}

\begin{eqnarray}\label{20}
\frac{dE}{dt}=-A\sigma_\kappa T^4,
\end{eqnarray}

\noindent where $\sigma_\kappa=\frac{J_3^\kappa(0)}{4\pi^2}$ in
which $J_3^\kappa(0)=\int_0^\infty\frac{x^3}{\exp_\kappa(x)-1}dx$
while $\exp_\kappa(x)=[\sqrt{1+\kappa^2x^2}+\kappa
x]^\frac{1}{\kappa}$, and we have
$\sigma_{\kappa\rightarrow0}=\sigma$ \cite{kanbb1}. Finally, we
reach

\begin{eqnarray}\label{21}
&&\!\!\!\!\!\!\!\tau_i^{\kappa}=-\frac{(8\pi)^3}{2\sigma_\kappa}\int_M^0m^2\cosh\big(4\kappa\pi m^2\big)dm,\nonumber\\
&&\!\!\!\!\!\!\!\tau_{ii}^{\kappa}=\frac{(8\pi)^3}{2\sigma_\kappa}\frac{M^3}{3},\\\nonumber
&&\!\!\!\!\!\!\!\tau_{iii}^{\kappa}=\tau_i^{\kappa},
\end{eqnarray}

\noindent for three cases we discussed above. In Fig. (\ref{fig6})
we have plotted $\tau_{ii}^{\kappa}$ and $\tau_{iii}^{\kappa}$
(family of black curves) for some values of $\kappa<0$ parameter
where we observe that the decay time grows as $\kappa$ tends to
larger values in negative direction. Such a behavior is parallel
to the satisfaction of the third law. For the second case we
observe that the decay time is finite for a finite mass black
hole.

\section{Summary}

Based on the third law of thermodynamics, it is impossible for a
system to touch zero-entropy state (at least a state with minimum
and finite value of entropy) as its temperature tends to zero by
only experiencing a finite number of thermodynamical processes. As
each process spends its own time interval to be completed, we may
even swap \textit{finite number of thermodynamical processes} with
\textit{finite time} in the above statement. The story becomes
more complicated in the ordinary black hole physics, generated by
Bekenstein entropy, where it is obtained that entropy diverges,
while its temperature approaches zero. In this situation, while
black hole evaporates at finite time~(\ref{5}), and loses all of
its mass, its temperature diverges at its final evolution steps
meaning that we face a catastrophe \cite{ali}. Here, we only
focused on Schwarzschild black hole as a primary solution which
clearly exposes the mentioned inconsistency with third law.

Motivated by the long range nature of gravity, some recent works
that propose a deep connection between quantum gravity and
generalized statistics \cite{epl,homa,mesri,barrow,mesri2}, and
also the successes of these types of statistics in justifying some
cosmological and gravitational phenomena
\cite{tsallis,refgerg,gerg,non13,nonK,EPJC,KHDE,sadeghi,mesri2},
we studied the status of third law for a Schwarzschild black hole
in the framework of some generalized statistics and we found out
that this law may theoretically be settled. Moreover, we obtained
that the thermodynamic analysis along with the laws may help us find new
energy definitions, thus establishing a consistency between the results
of thermodynamics and the predictions of quantum field theory
about the black body radiation. The latter means that
thermodynamics may eventually shed light on the physics and pave the way to find a
comprehensive energy definition \cite{energy,energy1}.

It is finally useful to mention that, a few months after
submitting our preprint to arXiv, we found two related papers
\cite{plb,gen1}. While one of them investigates the availability
of the generalized second law of thermodynamics in a universe
whose apparent horizon meets the generalized entropies \cite{plb},
another one studies the BH temperature and energy by employing the
Tsallis and Cirto and R\'{e}nyi entropies \cite{gen1}. The
approach of Ref.~\cite{gen1}, and additionally, its findings about
the Tsallis and Cirto entropy are similar to what we did and
obtained in Sec.~(\ref{TC00}), considered as a confirmation for
our concern and the strategy adopted here.

%%%%%%%%%%%%%%%%%%%%%%%%%%%%%%%%%%%%%%%%%%%%%%%%%%%%%%%%%%%
\acknowledgments{IPL would like to acknowledge the contribution of
the COST Action CA18108. IPL was partially supported by the
National Council for Scientific and Technological Development -
CNPq grant 306414/2020-1. J. P. M. G is supported by CNPq under
Grant No. 151701/2020-2.}
%%%%%%%%%%%%%%%%%%%%%%%%%%%%%%%%%%%%%%%%%%%%%%%%%%%%%%%%%%%%%%

\end{document}